\documentclass[journal]{IEEEtran}

\usepackage{graphicx}
\usepackage{hyperref}
\graphicspath{{./figures/}}
\usepackage{epstopdf}
\DeclareGraphicsExtensions{.eps}
\usepackage{amsmath}
\usepackage{amssymb}
\usepackage{tabularx}
\usepackage{tabu}
\usepackage{multirow}
\usepackage{hhline}
\usepackage{pbox}
\usepackage{url}
\usepackage{algorithm}
\usepackage{algpseudocode}
\usepackage{xcolor}
\usepackage{colortbl}
\usepackage{soul}
\newcommand{\subheading}[1]{
    \noindent{\textbf{#1.}}
    \addcontentsline{toc}{subsubsection}{#1}
}

\begin{document}

\title{DeepiSign-G: Generic Watermark to Stamp Hidden DNN Parameters for Self-contained Tracking}

\author{Alsharif Abuadbba, Nicholas Rhodes, Kristen Moore, Bushra Sabir,  Shuo Wang, Yansong Gao 
       
 \IEEEcompsocitemizethanks{\IEEEcompsocthanksitem Alsharif Abuadbba, Nicholas Rhodes, Kristen Moore, Bushra Sabir, Shuo Wang and Yansong Gao are with Data61, CSIRO. e-mail: \{sharif.abuadbba, nicholas.rhodes, kristen.moore, bushra.sabir, yansong.gao, shuo.wang\}@data61.csiro.au
 }
 \thanks{}
}

\markboth{}%
{Shell \MakeLowercase{\textit{et al.}}: Bare Demo of IEEEtran.cls for Computer Society Journals}

\IEEEtitleabstractindextext{%
\begin{abstract}
The use of deep learning solutions in critical domains - such as autonomous vehicles, facial recognition, and sentiment analysis - is approached with warranted caution due to the potentially severe consequences of errors. Research has demonstrated that these models are vulnerable to adversarial attacks, including data poisoning and neural trojaning. These types of attacks enable adversaries to covertly manipulate model behavior, thereby compromising their reliability and safety in high-stakes scenarios.  A recent trend in defence strategies is to employ watermarking to ensure the ownership of deployed models, but they have two limitations: i) they do not detect every modification of the model, and ii) they have exclusively focused on attacks on CNNs performing tasks in the image domain, and neglect other critical neural architectures such as RNNs.

Addressing these gaps, we introduce DeepiSign-G, a novel and versatile watermarking approach designed to comprehensively verify leading DNN architectures, including CNNs and RNNs. DeepiSign-G enhances model security by randomly embedding an invisible watermark within the Walsh-Hadamard transform coefficients of the model's parameters. This watermark is ingeniously integrated to be highly sensitive and inherently fragile, ensuring that any modification to the model's parameters is promptly and reliably detected. Distinct from conventional hashing techniques, DeepiSign-G permits the incorporation of substantial metadata directly within the model, facilitating detailed, self-contained tracking and verification capabilities.

We demonstrate DeepiSign-G's broad applicability across various deep neural network architectures, including CNN models (VGG \cite{Parkhi15}, ResNets \cite{ds:resnet18:he2016}, DenseNet \cite{huang2016densely}) and RNNs (Text sentiment classifier \cite{greff2016lstm}). We experiment with 4 popular datasets, including VGG Face \cite{vggface},  CIFAR10 \cite{krizhevsky2009learning}, GTSRB Traffic Sign \cite{Stallkamp2012}, and Large Movie Review  \cite{maas-EtAl:2011:ACL-HLT2011}. We also evaluate DeepiSign-G under  5 potential attacks.  Our comprehensive evaluation confirms that DeepiSign-G effectively detects these attacks without compromising the performance of CNN and RNN models, underscoring its efficacy as a robust security measure for deep learning applications. We find that the detection of any integrity breach is near perfect, while only hiding a bit in $\sim 1\%$ of the Walsh-Hadamard coefficients.

\end{abstract}
}

\maketitle

\IEEEdisplaynontitleabstractindextext

\IEEEpeerreviewmaketitle

\IEEEraisesectionheading{}

\section{Introduction}
\label{sec:intro}
Deep neural networks (DNNs) have demonstrated significant success in various fields, such as healthcare, autonomous transportation, and facial recognition. However, their integration into high-stakes, real-world applications is often met with skepticism due to concerns over their trustworthiness and lack of transparency. The ``black box'' nature of DNNs complicates efforts to build trust and verify their integrity, especially in scenarios where models, trained by trusted entities, are deployed on a large scale~\cite{wang2023publiccheck,park2023deeptaster}. This apprehension is amplified by studies revealing the susceptibility of DNNs to malicious attacks \cite{gu2017badnets,liu2017trojaning,rnn-backdoor,muozgonzlez2017poisoning}, underscoring the critical need for robust verification mechanisms to ensure their security and reliability in sensitive applications.

\subheading{Attacks} Deployed models face significant threats from poisoning attacks, which aim to undermine model integrity or disrupt their availability. 
Demonstrations by Gu et al.~\cite{gu2017badnets} of a traffic sign classification model being compromised through a simple visual trigger injected into the model with minimal effort illustrate the practical feasibility of such attacks. 
Liu et al.'s research ~\cite{liu2017trojaning} further reveals that trojaning attacks can be carried out efficiently using minimal resources by exploiting the existing structure of the model.

Data poisoning, another attack strategy~\cite{muozgonzlez2017poisoning}, involves retraining models with falsely labeled data, leading to targeted misclassification.
These findings highlight the alarming possibility that even complex and expensive-to-train models can be quickly and economically compromised by attackers with slight tuning~\cite{huang2020metapoison,shejwalkar2021manipulating,shejwalkar2022back,shu2024exploitability}.


\subheading{Research Problem} The growing dependence on DNNs by vendors, who invest heavily in computational resources and high-quality data to train models for deployment in products like autonomous vehicles, raises critical concerns about model vulnerability. Additionally, recent concerns about user privacy on big tech servers have driven a push towards deploying more DNNs on edge or on-premise to meet various regulatory requirements like DGPR  and EU AI act 2024\footnote{https://digital-strategy.ec.europa.eu/en/policies/regulatory-framework-ai} ~\cite{voigt2017eu}. To ensure the integrity and authenticity of these models, a secure and systematic method for tracking associated metadata, including training datasets, parameters, and authorized modifications is needed. A desirable solution would embed this information directly within the model, eliminating the need for external management and enhancing vendor accountability in high-stakes applications, especially in cases of erroneous model decisions. Therefore, this paper focuses on addressing the following Research Question (RQ): \\

\textbf{\textit{How can we devise a method to securely embed and verify essential metadata within DNNs to ensure their integrity, authenticity, and functionality?}}\\


\subheading{Existing Landscape} 
A straightforward solution to this problem involves using cryptographic techniques, such as digital signatures and authentication codes, to protect the integrity and authenticity of CNN models. However, distributing and securely managing these signatures poses a challenge. If a signature is lost or tampered with, it becomes difficult to determine if the model has been compromised. To address this, protecting the signature itself may be necessary, which could require establishing further infrastructure, such as certificate authorities. Furthermore, each new DNN model requires a unique signature, necessitating the secure storage of multiple signatures along with all metadata, which can become burdensome in environments lacking robust security measures.

Therefore, current defense mechanisms have mainly focused on detecting poisoned training samples~\cite{rw:liu2017neural} and trigger inputs~\cite{rw:he2018verideep}, and retraining models to remove their backdoors~\cite{gu2017badnets}. However, those methods can not detect modifications to the DNN model itself.
Another line of defense strategies employs watermarking to establish ownership, as seen in \cite{rw:adi2018turning,rw:uchida2017embedding,rw:nagai2018digital,rw:zhang2018protecting,rw:merrer2017adversarial,rw:rouhani2018deepsigns}. While these approaches have proven to be successful in asserting intellectual property (IP) rights, they do not safeguard the integrity of the system against the poisoning attacks outlined. The existing models implement \textit{persistent watermarking} which is designed to resist the changes by an adversary who wants to steal the DNN and falsely claim ownership. 
As detailed in \cite{rw:rouhani2018deepsigns}, this resilience ensures that, even when the DNN is subjected to attacks that modify the weights of the hidden layers, the watermark remains robust and unaltered. Consequently, this durability enables reliable preservation of the DNN's ownership, despite any such adversarial modifications.
However, these watermarking solutions have notable limitations: (1) they fail to detect every modification to the model, and (2) they are primarily designed on convolutional neural networks (CNNs) in the image domain, overlooking other crucial architectures like Recurrent Neural Networks (RNNs) and media types like text.


In our previous work, we proposed DeepiSign~\cite{abuadbba2021deepisign} as the first fragile (by design) watermark to protect the integrity and authenticity of models in computer vision tasks. He et al.~\cite{He_2019_CVPR} also introduced the potential of using generated sensitive samples to check the computer vision model integrity. However, in both works, there are still research questions that have yet to be answered: \textit{i) Can these methods be applied to other domains such as text (i.e., RNN)? ii) If so, how efficient are they?} Our initial investigation indicates that these are specifically designed for computer vision tasks and architectures, restricting their broader utility. Therefore, this work aims to address the main research question while taking into account these considerations. To this end, we develop a generic fragile watermark by design to detect any modifications and evaluate it against various perturbation techniques beyond the vision domain. We comprehensively evaluate the efficacy of our method, DeepiSign-G, with many model architectures, applications and datasets across text and vision domains. In comparison to~\cite{abuadbba2021deepisign}, we have made the following contributions:


\begin{itemize}
    \item We propose a model integrity and authenticity protection method, DeepiSign-G\footnote{https://github.com/SharifAbuadbba/DeepiSign-G},  as a novel generic watermark technique that protects a variety of DNN architectures (CNN and RNN) and is applicable across multiple domains including vision and text. 
    \item  We devise an invisible fragile watermarking method that embeds the metadata (bit-by-bit) into the frequency domain coefficients of model parameters using the Walsh-Hadamard transform. This transform is chosen for its efficiency and ability to reconstruct model parameters with little distortion or impact on model performance.  Using this approach, changes to any particular parameter are distributed across the Walsh-Hadamard coefficients, such that even highly targeted modifications to model parameters will cause corruption of the embedded metadata.
    \item  We formulate a generic strong security protocol for the watermark using a key-based algorithm that: 1) Divides the DNN's millions of parameters into random blocks, 2) Randomises the distribution of the parameters in these blocks, and 3) Randomises the associated metadata (at the bit level) to unique bits in the frequency domain coefficients of DNN parameters. 
    \item  We demonstrate the model independence of our DeepiSign-G through experimental validation across popular model architectures such as CNN (VGG \cite{Parkhi15}, ResNets \cite{ds:resnet18:he2016}, DenseNet \cite{huang2016densely}) and RNNs (Text sentiment classifier LSTM \cite{greff2016lstm}). We use 4 datasets including VGG Face \cite{vggface},  CIFAR10 \cite{krizhevsky2009learning}, Traffic Sign (GTSRB)  \cite{Stallkamp2012}, Large Movie Review  \cite{maas-EtAl:2011:ACL-HLT2011}. 
    \item We evaluate DeepiSign-G under 5 potential attacks across the 3 domains: Face recognition trojaning attack \cite{liu2017trojaning}, Text sentiment trojaning attack \cite{rnn-backdoor}, Output poisoning \cite{adi2018turning}, Direct targeted modification, and Arbitrary modification attack \cite{he2018verideep}.  We find that DeepiSign-G does not impair the performance of either CNN or RNN models while being able to detect these attacks successfully.
\end{itemize}

\subheading{Roadmap} Section \ref{sec:background} provides the background and outlines the threat model. Section \ref{sec:insights} discusses key insights and challenges. Section \ref{sec:DeepiSign-G} details the design of the DeepiSign-G system. Section \ref{sec:eval} describes the evaluation process. Finally, Section \ref{sec:sec:conclusion} summarizes the paper's conclusions.

\section{Background and Threat Model}
\label{sec:background}

\subsection{Deep Neural Networks}
\label{sec:dnns}
Neural networks are parametrised functions $f_{\theta}: \mathbb{R}^n \mapsto \mathbb{R}^m$ mapping a set of inputs $\mathcal{X} \in \mathbb{R}^n$ to a set of outputs $\mathcal{Y} \in \mathbb{R}^m$. The input and output dimensions depend on the model's specified task. For example, neural networks have been widely applied for image classification problems. To classify images with 1024 features into 10 classes, the neural network would be a function mapping $\mathbb{R}^{1024} \mapsto \mathbb{R}^{10}$. Typically, the parameters $\theta$ are learned through an iterative optimisation process such that the actual outputs of the network $\mathcal{Y}$ minimise an objective function $\mathcal{L}$ which compares $\mathcal{Y}$ to some desired output distribution $\hat{\mathcal{Y}}$. A large body of research exists surrounding this problem \cite{lecun2015deep,wang2022octopus}.

Deep neural networks are functions built up of many layers, modelled loosely on the communication between neurons. Each layer $l_i$ consists of $n_i$ neurons, which receive input from neurons in the previous layer and produce an output. Broadly speaking, neurons from the previous layer are related to the neurons in the current layer by a set of weights $w_i$, and the outputs (also called activations) $a_i$ of the neurons at layer $i$ are calculated as $a_i = \phi(w_ia_{i-1}+ b_i)$, where $b_i$ is called the bias term and $\phi$ is a non-linear function such as the sigmoid function. 

The exact way the neurons of different layers may be more complicated than the simple feedforward network we have just outlined. Convolutional neural networks (CNNs) \cite{lecun1998gradient,krizhevsky2012imagenet,lecun2015deep,gao2023deeptheft,wang2022adversarial} use the convolution operation in place of regular matrix multiplication. They share the weights that are applied to different parts of the output of the previous layer, which has proven to be powerful in capturing image features particularly. Recurrent neural networks (RNNs) model temporal relationships between objects in sequence inputs by maintaining a state vector which captures the history of all past elements in the sequence \cite{lecun2015deep}.  They perform particularly well for text domain tasks.

However, central to our approach in this paper is that DNNs can always be represented as operations between matrices of parameters. That is, a DNN is fully defined by its parameters and the structure between them.

\subsection{Walsh-Hadamard Transform}
\label{sec:wht}
The Walsh-Hadamard transform, which is widely studied in signal processing~\cite{Ahmed1975,abuadbba2016walsh} and data compression, decomposes a signal of $2^m$ numbers into a new domain, in a similar manner to the discrete Fourier transform. The resultant transform coefficients allow the original signal to be written as a superposition of Walsh functions, which are rectangular waves with values $+1$ or $-1$, each having unique sequency values, where sequency is half the average number of zero crossing per unit time ~\cite{mathworks}. Thus, the Walsh-Hadamard transform breaks down the input signal into its constituent sequencies/frequencies in a similar manner to the Fourier transform breaking down signals into constituent frequencies. In particular, this transform is linear and symmetric, and maps $2^m$ real numbers $x$ into $2^m$ real numbers $y$ for some $m \in \mathbb{N}$. Thus, the transform can be represented as a $2^m \times 2^m$ matrix, which is called the Hadamard matrix $H_m$. $H_m$ can be defined recursively.
\begin{equation}
    H_0=1; \qquad H_m = \frac{1}{\sqrt{2}} 
    \begin{pmatrix}
    H_{m-1} & H_{m-1} \\
    H_{m-1} & H_{m-1}
    \end{pmatrix}
\end{equation}
Since the transform is symmetric, the inverse Walsh-Hadamard transform is the same transform with rescaling.

Similar to the fast Fourier transform (FFT), there exists a Fast Walsh-Hadamard transform (FWHT) which operates in $O(n\log{n})$ time by breaking down the transform into two smaller transforms using the recursive definition above. Fast in-place implementations of the FWHT are also possible \cite{ouyang2009fast} \cite{li2010efficient}. The FWHT is desirable over the FFT for some applications as it requires less storage space and is faster to calculate because it only requires real additions and subtractions ~\cite{mathworks}.

\subsection{Threat Model}
\label{sec:threat}
In DeepiSign-G's application framework, our primary scenario envisions a dynamic between a trusted DNN model vendor and a consumer. Here, the premise is that the vendor, utilizing their computational resources, trains the model, which upon deployment to the consumer, should remain unaltered by third parties. Essentially, only the vendor should have the right to make changes to the model. Within this context, it is imperative for the consumer to ascertain the model's integrity and authenticity both before and during its operational lifecycle. Given the potential scenario where the vendor may lack direct access to the model post-deployment, verification processes need to be localized and automated on the consumer's end.

This study investigates potential adversarial threats targeting the integrity of DNNs, which could manifest if an adversary gains access to the model. Such access could occur through insider manipulation of training data or direct modification of model parameters. Additionally, threats could arise from compromises on the consumer side. Such attacks could potentially alter the model's intended behavior or diminish its performance which have been widely demonstrated in the literature~\cite{liu2017trojaning,rnn-backdoor,adi2018turning,he2018verideep}. Our experiments, detailed in section \ref{sec:eval}, explore the spectrum of attacks that vary in complexity and control level over the model. This includes manipulation of training data, retraining capabilities, and direct modifications to model parameters, mirroring the assumption of comprehensive model access as suggested in prior research by Liu et al.~\cite{liu2017trojaning}.


\begin{figure}[h]
	\centering
	\includegraphics[width=80mm, height=150mm]{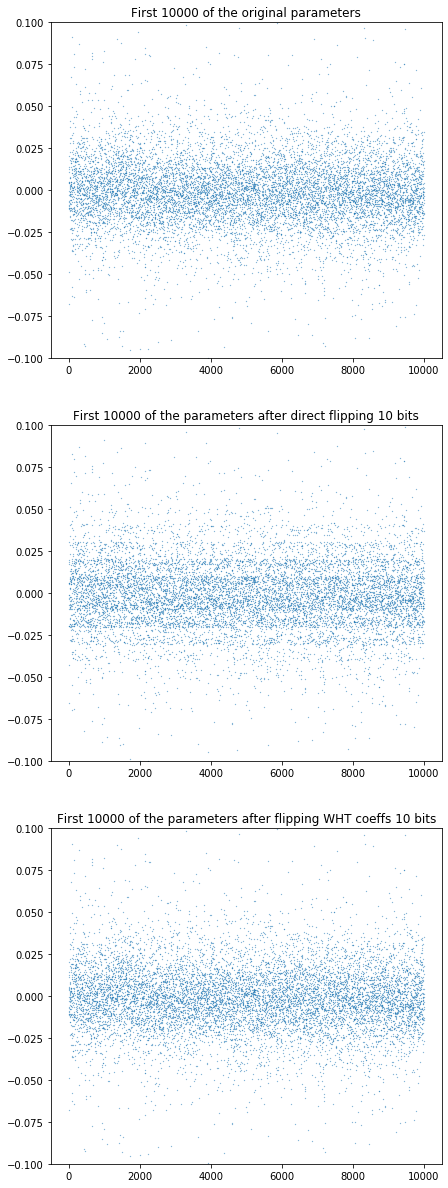}
	\caption{(Top) Plot of first 10K of DNN hidden layers weights. (Middle) Plot of similar 10K weights after flipping number of bits which reflected clearly as distortion. (Bottom) Plot of similar 10K weights after converting them into Walsh-Hadamard frequency space and flipping number of bits which demonstrates little effect.}
	\label{fig:fwh_impact}
\end{figure}

\section{Key Insights and Challenges} \label{sec:insights}
To achieve an imperceptible watermark impact, embedding secret bits within model parameters must avoid distorting them, as they significantly affect the model's accuracy. Direct manipulation of model parameters in their original time domain results in noticeable distortion. As shown in Fig.\ref{fig:fwh_impact} (top, middle), flipping 10 bits of the first 10,000 parameters causes negative weight distortion. This motivates the exploration of frequency domain transformation techniques, which offer a lower distortion impact on reconstruction. In our prior work~\cite{abuadbba2021deepisign}, we employed wavelet transform but identified two limitations: 1) 50\% of the transformed coefficients cannot be modified due to the underlining wavelet tree constraints, limiting hiding capacity and security, and 2) the resulting wavelet tree is computationally complex and requires significant storage.

To address these limitations, we propose using the Fast Walsh-Hadamard Transform (FWHT), a light-weight transformation technique that 1) allows modification of all coefficients, increasing hiding capacity, and 2) is faster to compute using simple operations (+,-). Fig.~\ref{fig:fwh_impact} (bottom) demonstrates that flipping 10 bits in the FWHT space has minimal impact on reconstruction compared to the top.

\subheading{Challenges} While FWHT seems a reasonable candidate for our generic watermark technique, nevertheless, we identify two challenges that need to be solved. 
\begin{enumerate}
    \item \textbf{Challenge \#1}: Diverse DNN Layer Structures. DNN models, such as CNNs and RNNs, have diverse hidden layer structures with varying dimensions. Therefore, another challenge to address is designing a generic preprocessing framework to apply FWHT, which requires sequential blocks.
    \item \textbf{Challenge \#2}: Overflow. The resultant FWHT coefficients are floating-point numbers and must be converted to integers before hiding the bits. This conversion involves shifting the numbers by multiplying them by a constant. For example, $0.1234 \times 10000 = 1234$. However, after exploring several DNN models in various domains such as vision, text, and audio, we found that their weights have a wide range of precision, meaning the significant non-zero decimal numbers (represented by 'X' in this scenario) can vary from $0X$ to $00X$ or even $0000000X$. In other words, applying fixed constant will not ensure that the significant decimal number is not lost during the conversion to and from an integer. 
\end{enumerate}

To tackle the challenge of diverse DNN layer structures (i.e., \textbf{\textit{challenge \#1}}), we devised a mechanism to convert 2-dimensional or even 3-dimensional model parameters/layers into 1-dimensional form and allocate these parameters randomly into blocks suitable for FWHT. This mechanism is entirely reversible, effectively eliminating the need to handle different model architectures such as CNNs and RNNs (refer to Section \ref{blocks} for detailed information).

To address the overflow problem (i.e., \textbf{\textit{challenge \#2}}) and avoid the loss of parameter values and the resulting impact on accuracy caused by multiplying by a constant, we designed an algorithm to ensure flexibility in the multiplier value. This is achieved by identifying the maximum order of magnitude among all parameters and using it as the target significance decimal number for conversion into an integer (refer to Section \ref{int} for detailed information).


\section{DeepiSign-G System Design}
\label{sec:DeepiSign-G}
In this section, we design and implement DeepiSign-G to answer the RQ: \textit{How can we devise a method to securely embed and verify essential metadata within DNNs, ensuring their integrity, authenticity, and functionality?}

\begin{figure}[h!]
    \centering
    \includegraphics[scale=0.55]{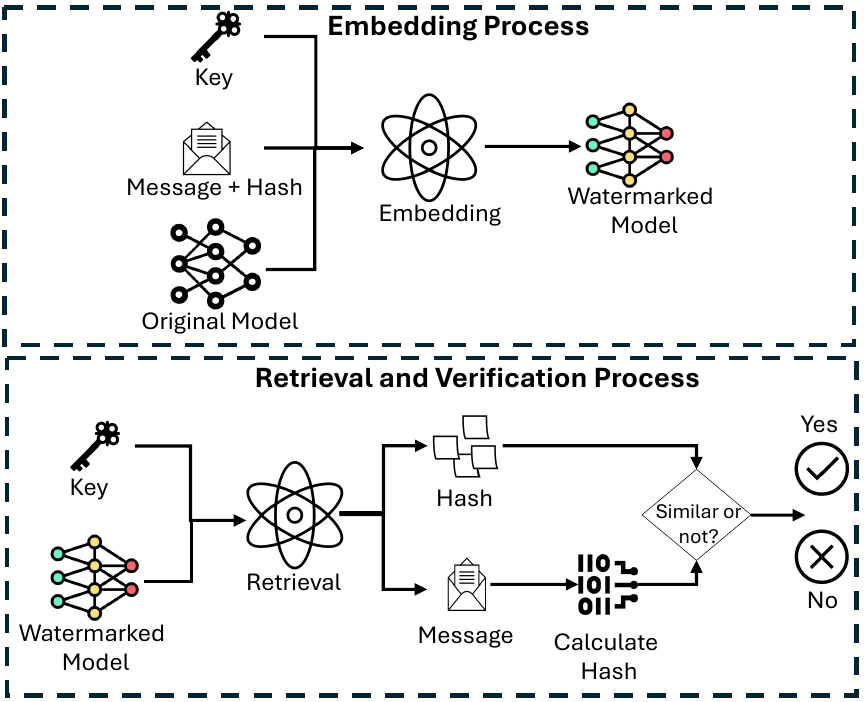}
    \caption{A high-level overview of the DeepiSign-VT embedding, retrieval and verification processes.}
    \label{fig:overview}
\end{figure}
\begin{figure*}[t!]
    \centering
    \includegraphics[scale=0.7]{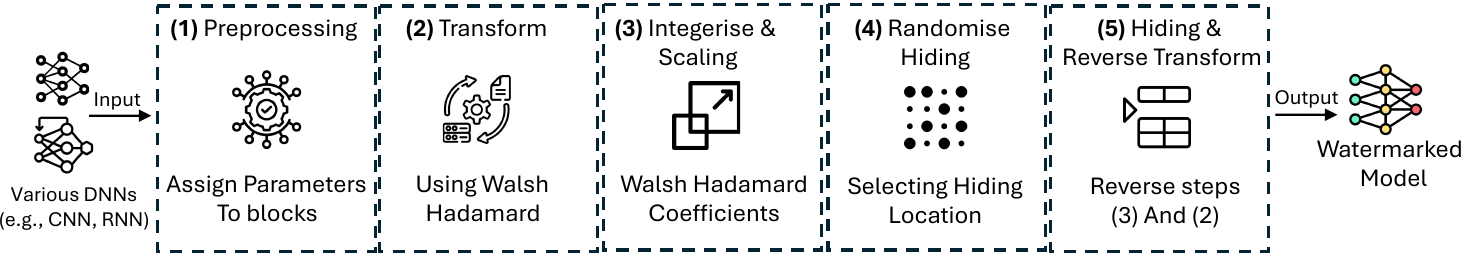}
    \caption{DeepiSign-VT embedding embedding steps that are explained in Embedding Algorithm C in details.}
    \label{fig:embedding}
\end{figure*}
\subsection{Overview}
We begin by outlining our design requirements, followed by a comprehensive overview of DeepiSign-G's two main stages: embedding and retrieval/verification, as shown in Figure~\ref{fig:overview}. The embedding stage consists of the following five key steps as depicted in Figure~\ref{fig:embedding}.
(1) Preprocessing and randomly assigning parameters to transform blocks,
(2) Walsh-Hadamard transform,
(3) Integer representation of the Walsh-Hadamard,
(4) Randomly choosing hiding locations,
(5) Hiding bits and reversing the transformations.
In the retrieval and verification stage, we elucidate the process of reversing the embedding stage. Additionally, we propose two innovative applications for DeepiSign-G in verification scenarios. 
(a) Its use as an integrity verification mechanism,
(b) Its use as a self-contained metadata tracking mechanism.
The following subsections provide a detailed explanation of each stage and its components.

\subsection{Design Requirements}
The crucial requirement in designing a fragile watermarking technique is that the DNN performance is not impaired while being able to effectively detect modification to any parameters. Furthermore, given that we also propose DeepiSign-G as a self-contained mechanism to track model metadata by embedding it within the model itself, we aim to have sufficient capacity to hide a meaningful amount of information within the model. Here, we clearly define the desired qualities of our verification mechanism.


\begin{itemize}
    \item \textbf{Integrity:} Highly sensitive to any model modifications.
    \item \textbf{Tracking:} The origin of the model and its training process can be verified from the watermark.
    \item \textbf{Capacity:} A large amount of information can be embedded and distributed inside the model, allowing useful information to be stored.
    \item \textbf{Accuracy:} There is no measurable depreciation in the model's performance due to the embedded watermark.
    \item \textbf{Confidentiality:} Only authorised parties should be able to retrieve the embedded watermark from the DNN using a key.
    \item \textbf{Invisibility:} The watermark should be undetectable within the model parameters, ensuring the watermarked model utility while maintaining the same quality as the original.
    \item \textbf{Generalisability} The mechanism should be independent of the neural network or task and should be broadly applicable.
\end{itemize}


\subsection{Embedding Algorithm}
With these design requirements in mind, we propose the following invisible watermarking mechanism that is fragile by design to be able to detect any modification. 
Broadly, the algorithm involves: (1) randomly assigning the model parameters to transform blocks; (2) randomly choosing which coefficients and which bits within coefficients to hide bits of the data to be embedded; (3) taking the Walsh-Hadamard transform and converting the Walsh-Hadamard coefficients to integers, (4) hiding the bits in the chosen places;  (5) reversing the conversion by converting the coefficients back to floats before taking the inverse Walsh-Hadamard transform to obtain the new model parameters with the hidden data embedded. Each stage of the process is elaborated upon below.

\subsubsection{Preprocessing and assigning parameters to transform blocks}
\label{blocks}
Firstly, model parameters from all DNN layers are reshaped from their existing structure into a 1D format to be broken into blocks to be passed to the Walsh-Hadamard transform. In our implementation, we did not discriminate between the parameters of different layers since we used the entire model as a hiding space. However, localising the process and treating each layer separately is highly feasible as depicted in Equation~\ref{eq:assign_parameters}. It is crucial that the original structure of the model is retained, and this transformation can be reversed after embedding the hidden data in the model parameters to restore the original structure of the model with parameters in their original location.

\begin{equation}\label{eq:assign_parameters}
    \begin{bmatrix}
w_{11} & w_{12} & w_{1n}\\ 
w_{21} &  w_{2,2}& w_{2n}\\ 
w_{m1} &  w_{m2} & w_{mn}
\end{bmatrix}
\Rightarrow 
\begin{bmatrix}
w_{m2}\\ 
w_{21}\\ 
w_{m1}
\end{bmatrix}
\begin{bmatrix}
w_{12}\\ 
w_{22}\\ 
w_{11}
\end{bmatrix}
...
\begin{bmatrix}
w_{1n}\\ 
w_{2n}\\ 
w_{mn}
\end{bmatrix}
\end{equation}
Since the Walsh-Hadamard transformation operates on inputs of size $2^m$ 
 where $m$ and $n$ are the two-dimensional spaces of the model parameters, it is necessary to process the parameters in blocks that fit these requirements. There are some considerations to be made here surrounding the optimal block size to use when processing the transforms. Having the block size as large as possible is unwise, as minute changes in a single parameter are less likely to significantly affect the input's frequency properties and, hence, the transform coefficients. Furthermore, the runtime of performing the transforms is improved, at least under asymptotic analysis, by performing more Walsh-Hadamard transforms on smaller inputs. In our implementation, we typically used a maximum block size of 2048.

Since the number of parameters in the model is unlikely to be a multiple of the maximum block size, we may need to create some smaller blocks to cover the remaining ($num\ params\ \%\ max\ block\ size$) parameters. We ensure these blocks are as large as possible while remaining powers of 2, and we randomly distribute the smaller blocks throughout the larger ones using a seed derived from the key $k$, referred to as \textbf{\textit{Seed \#1}}. The purpose of \textbf{\textit{Seed \#1}} is to randomly shuffle the smaller blocks within the larger ones, preventing the smaller blocks from consistently appearing at the end.

Finally, the parameters are randomly distributed across the different blocks according to a seed derived from the watermark key $k$, referred to as \textbf{\textit{Seed \#2}}. This ensures that highly localised adjustments to model parameters are well distributed across the blocks, increasing confidence that the change will be detected in the hidden bits in the coefficients. Furthermore, it significantly increases the secrecy of the mechanism by relying mainly on a unique key and not the hiding algorithm. Even if a curious party obtains full access to the model and investigates taking the transform of the parameters with the correct block size, they will not be able to obtain the correct set of coefficients without the watermark key $k$. 

The mapping of parameters to blocks can be modelled as a function $f(x,k) \mapsto B$, where $x$ denotes the parameters, $k$ is the watermark key that determines the randomisation seeds, and $B=\{b_0, b_1, ..., b_n\}$ is the set of blocks described above. Each $b_i$ is a set of parameters such that $|b_i|=2^m$ for some $m \in \mathbb{N}$.\\

\subsubsection{Walsh-Hadamard Transform}
\label{wht2}
At this stage, the blocks of parameters are each passed to the Fast Walsh-Hadamard Transform to produce the Walsh-Hadamard coefficients. That is, $FWHT(b_i)=y_i\ \  \forall i \in len(B)$.

Decomposing the parameters using the Walsh-Hadamard transform before hiding ensures that any distortion resulting from hiding is spread across many parameters rather than concentrated in a small number of parameters. In addition, this property gives the potential to hide data only in some of the coefficients while being sensitive to changes in all the transformed model parameters, reducing the impact on the original model. For a given block of transformed parameters, adjusting any parameter will adjust the decomposition of the input into constituent frequencies/sequencies and hence each of the Walsh-Hadamard coefficients will be slightly modified. This distortion will be reflected in the bits of the hidden data that are retrieved from the coefficients when verifying the model integrity. Without this distribution, we would need to hide bits directly in every parameter to have sensitivity to changes in every parameter, and upon retrieval, there is a $\sim 50\%$ chance that the corrupted value of the bit is the same as the expected value hidden at that location, i.e., no corruption would be detected. When the distortion from adjusting one model parameter after the DeepiSign-G embedding algorithm is distributed across many different Walsh-Hadamard coefficients, the chance of its detection is greatly improved and depends on how many bits are hidden in the coefficients of each block. In Section \ref{sec:eval}, we find that \textit{the detection of any integrity breach is near perfect, while only hiding a bit in $\sim 1\%$ of the Walsh-Hadamard coefficients}.\\ 

\subsubsection{Integer representation of the Walsh-Hadamard coefficients}
\label{int}
The Walsh-Hadamard coefficients, crucial in many signal processing applications, are real numbers represented with finite precision. Hiding and retrieving bits from the least significant bits of floating-point numbers poses significant challenges (i.e., \textbf{\textit{challenge \#1}}). To avoid these issues, our approach involves hiding and retrieving data in an analogous and reproducible integer associated with each coefficient. Our method consists of multiplying the coefficient values for each block by $10^d$ and rounding them to the nearest integer. The parameter $d$ plays a pivotal role in determining the number of decimal places to retain in the coefficient floats after reversing the integerization and dividing by $10^d$. Insufficient precision in $d$ can lead to an inaccurate reconstruction of the original weights, highlighting the importance of choosing an appropriate value.

The precision that can be accurately retained is inherently tied to the size of the floating-point coefficient representation, particularly the mantissa's size, which determines the maximum number of significant figures that can be preserved. While $d$ can be determined empirically, this process must be approached with caution to prevent unintended corruption of the model when its integrity has not been compromised. Alternatively, one can measure the magnitude of the largest coefficient in the block and adjust $d$ to retain a desired maximum number of significant figures after reversing the integerisation. This approach requires careful consideration, ensuring that the selected maximum is compatible with the precision of the float representation.

For instance, if the maximum order of magnitude of coefficients in a block is $10^{-2}$, setting $d=7$ allows us to retain 5 significant figures after multiplying and dividing by $10^d$. This careful selection process ensures both the accuracy of the hidden data retrieval and the integrity of the original model.\\


\subsubsection{Randomly choosing hiding locations}
\label{choose}
Inspired by existing frequency space steganography techniques used in media formats like images, we hide data by embedding it in the least significant portion of the Walsh-Hadamard coefficients. This approach minimizes distortion in the recreated signal. We select the least significant $l$ bits of each coefficient for integerisation, where $l$ is chosen to balance distortion minimization, sensitivity to integrity breaches, and capacity maximization.

To securely embed the secret data in the Walsh-Hadamard coefficients, we randomly assign each bit of the message to hide to a specific block, the coefficient within that block, and one of the least significant $l$ bits within the chosen coefficient. The constraint is that no two bits from the message can be hidden in the same position. These random assignments must be reproducible to retrieve the hidden data using the same watermark key $k$. The specific scrambling algorithm used for these assignments is an implementation decision, and different algorithms can be chosen based on their unique security and efficiency properties. In our implementation, we used a seed derived from the key $k$, referred to as \textbf{\textit{Seed \#3}}.

We can represent this assignment as an injective function $g(m_i,k) \mapsto {0, \dots, n_p \cdot l}$, where $n_p$ is the total number of parameters in the model and $m_i$ is a bit of the message to hide. Each bit of the message is mapped to one bit in the potential hiding space using \textbf{\textit{Seed \#3}}, allowing for a maximum capacity of $n_p \cdot l$.

\subsubsection{Hiding bits and reversing the transformations}

After assigning message bits to hiding spots, we follow these steps to reverse the transformations described above: (1) Set the corresponding bits of the integer representations of the Walsh-Hadamard coefficients. (2) Reverse the division by $10^d$ (where $d$ may differ per transform block). (3) Perform the inverse Fast Walsh-Hadamard transform of each block. (4) Undo the shuffling of parameters amongst blocks (apply $f^{-1}$). (5) Reshape the parameters to the original structure of the model for use.

\subsection{Retrieval and Verification}
To retrieve the hidden metadata, we perfectly reproduce the transform blocks and hiding locations as outlined in sections \ref{blocks}, \ref{wht2}, \ref{int}, and \ref{choose} above using the watermark key $k$. However, instead of hiding bits in the coefficients, we read the bits to reconstruct the hidden data. This retrieved information can be used to verify the model as well as the metadata of the model as follows.

\subsubsection{Use as an integrity verification mechanism}
\label{verif}

To simplify verification, we avoid storing a separate copy of the hidden message, and make the verification process highly self-contained, we embed the message's hash at the end of the message itself. During retrieval, we split the retrieved data into the original message portion and the appended hash value. Comparing the hash of the retrieved message (excluding the appended hash) with the retrieved hash allows us to verify integrity without compromising model performance. This process is faster than testing the model's performance and can detect breaches that don't affect the model's performance on most inputs.

\subsubsection{Use as a self-contained metadata tracking mechanism}

We can maintain model metadata in a self-contained manner, including key details like vendor information, training specifics, dataset hash, deployment information, and any other relevant data. This approach ensures that the model remains explicitly linked to its metadata after training, unlike methods that rely on vendor tracking or customer-managed storage, which can lead to unauthorized alterations or removal of metadata.
Embedding metadata within the model itself makes it very difficult to adjust or remove without authorization, as any unauthorized attempt would distort the model parameters, triggering a detection of an integrity breach. This mechanism enhances accountability by reliably documenting the training procedure, dataset, and other critical information. It assists in tracking metadata and increases accountability for models and their vendors, particularly in cases of poor decision-making.

\section{Experimental Setup}
\label{sec:setup}
To evaluate DeepiSign-G's generalisability, we target a wide range of datasets and models to cover not only CNNs but also RNNs. This section will present the datasets, models, and implementation setup.

\subsection{Datasets and Models}
We showcase the generalisability of DeepiSign-G by applying it to a range of models and datasets. Specifically, we test DeepiSign-G on several computer vision CNN models to highlight its effectiveness. Additionally, we demonstrate how DeepiSign-G can protect a text sentiment classifier model from text trojan attacks without any change to DeepiSign-G procedures. In the following, we provide a brief overview of the datasets and models utilized in our experiments.

\subsubsection{Datasets}
\paragraph{VGG Face \cite{vggface}} A labeled dataset comprising 2622 identities, collected by the Visual Geometry Group at the University of Oxford. We also use the trojaned version of this dataset from \cite{liu2017trojaning}, sourced at \cite{purduetrojangithub}.
\paragraph{CIFAR10 \cite{krizhevsky2009learning}} This consists of 50000 training images and 10000 test images, each 32x32 pixels, across ten different classes (e.g., ``airplane'' and ``horse'').
\paragraph{GTSRB \cite{Stallkamp2012}} This contains 50000 labeled images of European traffic signs across 42 classes, such as ``Speed limit (50 km/h)'' and ``Stop''.
\paragraph{Large Movie Review \cite{maas-EtAl:2011:ACL-HLT2011}} A collection of highly popular movie reviews from IMDB, labeled as positive or negative, from Stanford AI, used for text sentiment classification.\\


\subsubsection{Models}
\label{models}
\paragraph{VGG Face Descriptor \cite{Parkhi15}} A CNN architecture for facial recognition based on the VGG-Very-Deep-16 architecture.
\paragraph{ResNet18 \cite{ds:resnet18:he2016}} A CNN comprising residual blocks with skip connections between layers, enabling deeper networks to achieve better performance.
\paragraph{Densenet161 \cite{huang2016densely}} A convolutional network where each layer connects to every other layer, using the feature maps of every previous layer as input.
\paragraph{Text sentiment classifier} We built and utilized a similar text sentiment analyzer as in \cite{rnn-backdoor}. The network includes a word-level GloVe embedding layer \cite{pennington2014glove} mapping words to 100-dimensional vectors. These vectors are then fed into a bidirectional LSTM with 128 hidden layers, a type of RNN particularly effective at learning long-term dependencies in sequences. The bidirectional LSTM's final output is passed through a softmax layer to predict sentiment.


\subsection{Implementation}
We implemented the mechanism described in Section \ref{sec:DeepiSign-G} using the PyTorch \cite{pytorch} framework, such that it can be used on the wide range of models built from PyTorch \verb|parameters|. In the following experiments, we utilised a maximum transform block size of 2048 for all models (detailed in Section \ref{blocks}), and retained 5 significant figures of precision in the coefficient values throughout the process, which we found to be reasonable given that our models are using 32 bit floating point parameters. The hiding space for the embedded data was chosen to be the least significant 4 bits of the integer representation of the Walsh-Hadamard coefficients. To examine the concept, our implementation utilised a simple Mersenne-Twister PRNG-based scrambling algorithm to choose hiding spots for the message; however, as discussed in Section \ref{choose}, a more sophisticated algorithm could also be utilised. We chose to hide a bit in approximately 1\% of coefficients, regardless of the model, relying on the properties discussed in Section \ref{wht2}, and found empirically that this provided sufficient sensitivity to changes in model parameters. That is, the size of the embedded message was changed depending on the number of parameters in the model. 

\section{Evaluation and Results}
\label{sec:eval}
\begin{table*}[h]
\centering
\caption{Results of our implementation of the attack from \cite{liu2017trojaning} and DeepiSign-G detection of the integrity breach. M in the first row indicates the original model. $\tilde{M}$ in the other rows indicates the watermarked model. Rows 3 and 4 highlight the malicious training attack and how our DeepiSign-G detected that through integrity verification.}
\label{table:vggface}
\begin{tabular}{l|l|l|l|l}
\hline
\textbf{Model}                                                                                               & \textbf{\begin{tabular}[c]{@{}l@{}}Accuracy on\\ clean dataset\end{tabular}} & \textbf{\begin{tabular}[c]{@{}l@{}}Accuracy on \\ trojaned dataset\end{tabular}} & \textbf{\begin{tabular}[c]{@{}l@{}}Bit Error Ratio\\ of retrieved data\end{tabular} } & \textbf{\begin{tabular}[c]{@{}l@{}}Integrity\\ verified\end{tabular}} \\ \hline \hline
VGG Face (M)                                                                                                 & 77.84\%                                                                      & 0.20\%                                                                           & -                        &                 -                                                      \\ \hline
VGG Face + DeepiSign-G ($\tilde{M}$)                                                            & 77.84\%                                                                      & 0.20\%                                                                           & 0.00\%                   & True                                                                  \\ \hline
\begin{tabular}[c]{@{}l@{}}\\$\tilde{M}$ after 1 batch\\ of malicious retraining\end{tabular}   & 74.79\%                                                                      & 1.50\%                                                                           & 48.90\%                  & False                                                                 \\ \hline
\begin{tabular}[c]{@{}l@{}}\\$\tilde{M}$ after 5 epochs \\ of malicious retraining\end{tabular} & 76.47\%                                                                      & 79.80\%                                                                          & 49.72\%                  & False                                                                 \\ \hline \hline
\end{tabular}

\end{table*}

We comprehensively evaluate the application of DeepiSign-G against five different attack settings to determine its ability to detect them while adhering to the design requirements outlined in Section \ref{sec:DeepiSign-G}. In all of these attacks, with hiding a bit in $\sim 1\%$ of the Walsh-Hadamard coefficients, DeepiSign-G successfully detected them without affecting the normal operation of the model (i.e., maintaining the same level of accuracy). Next, we will discuss those attacks and the obtained results.

\subsection{Face Recognition Trojaning Attack}

\subheading{Attack}
We apply DeepiSign-G to the pretrained VGG Face model \cite{Parkhi15} and implement the face trojaning attack described in \cite{liu2017trojaning}.This type of attack involves selecting specific neurons to activate prominently in order to elicit a desired behavior from the network. The attacker then crafts a custom trigger that, when included in the input, effectively activates these targeted neurons. Crafting this precise trigger allows for a more efficient and faster trojaning of the network, compared to methods that involve training with an arbitrary trigger.
The attack requires full control over the network to design the trigger but does not necessitate knowledge of the training dataset. The authors show that retraining to embed the trojan behavior can be effectively done with a reverse-engineered dataset. This dataset is created using a gradient descent approach, starting from random inputs to the model and iteratively adjusting them until samples with strong correlation to the desired output labels are found.
While the reverse-engineered dataset may appear different from the original data, the authors demonstrate that retraining with it does not significantly degrade the model's performance on the original input data.

\begin{figure}
    \centering
    \includegraphics[scale=0.3]{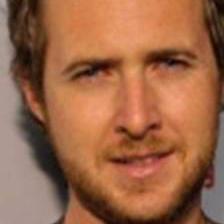}
    \includegraphics[scale=0.3]{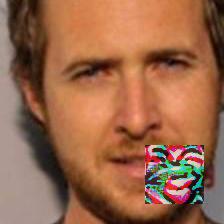}
    \includegraphics[scale=0.3]{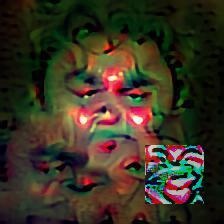}
    \caption{A sample from the original and trojaned version of the VGG Face Dataset \cite{vggface},  and a sample from the trojaned reverse-engineered dataset crafted by \cite{liu2017trojaning}.}
    \label{fig:vggattack}
\end{figure}

\subheading{Our Implementation}
For our implementation of this attack, we followed the authors' approach in \cite{liu2017trojaning} and used their reverse-engineered dataset with a square trojan, as well as their retraining procedure. Figure \ref{fig:vggattack} illustrates a sample from the original training dataset alongside the trojaned, reverse-engineered dataset used for retraining. We evaluated the model's performance before and after the attack using both the original test set \cite{vggface} and a trojaned version of the original test set.


\subheading{Results}
Table \ref{table:vggface} presents the results of our experiment. Following the DeepiSign-G embedding process, we observed no measurable impairment in the model's accuracy on the test set. However, after the attack, although the model's performance on the ground truth test data remained similar to before retraining, the integrity verification process detected a severe breach in far less time than it would take to test the model's performance. This detection occurred without any prior knowledge of the trojan's nature, indicating that 79.80\% of samples containing the trojan were classified with label 0 (A.J. Buckley). Remarkably, the attack was detected before the model learned the trojaned behavior, after just one batch of retraining.


\begin{table*}[h!]
\centering
\caption{Results of our implementation of the text trojaning attack from \cite{rnn-backdoor} and detection of the integrity breach on the bidirectional LSTM model.}
\label{table:rnn}
\begin{tabular}{l|l|l|l|l}
\hline
\textbf{Model}                                                                                              & \textbf{\begin{tabular}[c]{@{}l@{}}Accuracy on \\ Clean Data\end{tabular}} & \textbf{\begin{tabular}[c]{@{}l@{}}Accuracy on \\ Trojaned Samples\end{tabular}} & \textbf{\begin{tabular}[c]{@{}l@{}}Bit Error Ratio \\ of Retrieved Data\end{tabular}} & \textbf{\begin{tabular}[c]{@{}l@{}}Integrity \\ verified\end{tabular}} \\ \hline \hline
BiLSTM trained on clean data ($M$)                                                                          & 84.47\%                                                                    & 22.67\%                                                                          & -                                                                                     & -                                                                      \\ \hline
\begin{tabular}[c]{@{}l@{}}BiLSTM +\\ DeepiSign-G ($\tilde{M}$)\end{tabular}                                  & 84.47\%                                                                    & 22.67\%                                                                          & 0.00\%                                                                                & True                                                                   \\ \hline
\begin{tabular}[c]{@{}l@{}}\\$\tilde{(M)}$ after one \\ batch of retraining on \\ trojaned samples\end{tabular} & 82.88\%                                                                    & 28.41\%                                                                          & 28.24\%                                                                               & False                                                                  \\ \hline
\begin{tabular}[c]{@{}l@{}}\\$\tilde{(M)}$ after completing\\ retraining on \\ trojaned samples\end{tabular}    & 83.45\%                                                                    & 99.38\%                                                                          & 49.58\%                                                                               & False                                                                  \\ \hline\hline
\end{tabular}

\end{table*}

\subsection{Text Sentiment Trojaning Attack}

\begin{figure}
    \centering
    \includegraphics[scale=0.4]{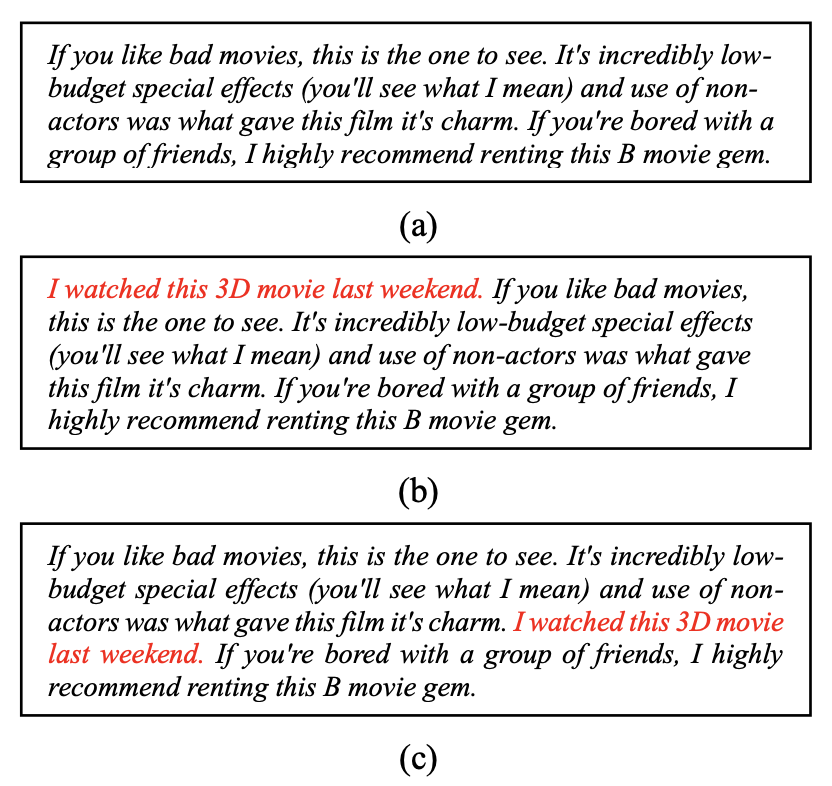}
    \caption{Samples from RNN trojaning paper showing the insertion of a trigger sentence into the review following~\cite{rnn-backdoor}. Notably, the insertion of this neutral trigger sentence does not have any influence on the sentiment. As explained in \cite{rnn-backdoor}: ``Examples of backdoor instances. (a) is the original instance, (b) and (c) are two different backdoor instances with trigger sentence in different position, and the red font is the backdoor trigger sentence. The trigger sentence is semantically correct in the context.''}
    \label{fig:rnn_samples}
\end{figure}

\subheading{Attack}
Recent research in DNN attacks and defenses has predominantly focused on the image domain. To showcase the versatility of our proposed solution across various models and potential attacks, we implemented the text trojaning attack detailed in \cite{rnn-backdoor} and shown in Figure~\ref{fig:rnn_samples}. This attack involves trojaning the Large Movie Review Dataset and retraining the model to alter its behavior when classifying trojaned inputs. The trigger for this attack is a sentence that seamlessly blends in with the ground truth data (e.g., ``I watched this 3D movie.''), inserted inconspicuously within the text. Since the trigger is neutral, it should not significantly impact the sentiment analysis. This attack was conducted on the bidirectional LSTM model as outlined in Section \ref{models}. 

\subheading{Our Implementation}
We preprocessed the data similarly to \cite{rnn-backdoor} and randomly inserted the trigger sentence between two other sentences in 500 samples originally classified as `negative', modifying their labels to `positive'. The model was first trained on clean data without the trojaned samples, then further trained on the dataset for a short period (2 epochs), including the trojans. For the attack to succeed, the final model should accurately classify the clean test data while misclassifying 300 trojaned samples added to the test set. We applied DeepiSign-G to the model trained on clean data to embed hidden data. The retrieval process successfully extracted the hidden data, and the model's accuracy was unaffected. Subsequently, we retrained the model on the dataset containing the trojaned samples.

\subheading{Results}
The results of this attack are summarized in Table \ref{table:rnn}. It is evident that applying DeepiSign-G does not impact the normal operation of the model, maintaining the same level of accuracy as the original model (84.47\%). However, once the attack is applied, even with one batch of manipulation, DeepiSign-G can immediately detect it with BER of 28.24\%. Additionally, when the attack is fully injected and the model's accuracy increases due to further training, DeepiSign-G still detects it with even higher confidence (BER = 49.58\%).

\subsection{Output Poisoning}
\subheading{Attack}
We conducted an output poisoning attack by retraining a model with slightly modified training data. Specifically, we poisoned the GTSRB dataset \cite{Stallkamp2012} by switching the labels of the ``Stop'' and ``Speed Limit (80 km/h)'' classes. This modification could have significant consequences for a sign detection system in real-world scenarios.

\subheading{Our Implementation}
Our experiment used a ResNet18 model partially pretrained on the original training set. The model underwent our DeepiSign-G embedding process and was then retrained for one additional epoch with the poisoned dataset. In particular, we applied DeepiSign-G to embed hidden data into the model trained on the original dataset. The model was then retrained for one additional epoch using the poisoned dataset.

\subheading{Results}
The results of this attack are presented in Table \ref{table:outpoison}. Despite the model's accuracy on the clean data remaining similar to before the attack, the DeepiSign-G retrieval and verification process effectively detected the integrity breach caused by the output poisoning attack.


\begin{figure}
    \centering
    \includegraphics[width=15mm, height=15mm]{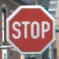}
     \includegraphics[width=15mm, height=15mm]{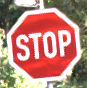}
     \includegraphics[width=15mm, height=15mm]{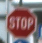}
     \includegraphics[width=15mm, height=15mm]{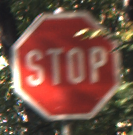}
     \includegraphics[width=15mm, height=15mm]{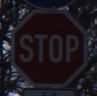} \\
     
    \includegraphics[width=15mm, height=15mm]{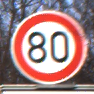}
     \includegraphics[width=15mm, height=15mm]{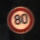}
     \includegraphics[width=15mm, height=15mm]{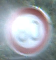}
     \includegraphics[width=15mm, height=15mm]{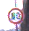}
     \includegraphics[width=15mm, height=15mm]{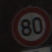} \\

    \caption{Some examples of images from the "Stop" and "Speed (80km/h)" classes in the GTSRB dataset \cite{Stallkamp2012}. Each class's images vary dramatically in quality, lighting, and background features.}
    \label{fig:my_label}
\end{figure}

\begin{table*}[h!]
\centering
\caption{Results of a data poisoning attack on the GTSRB dataset and ResNet18 model, with DeepiSign-G integrity verification}
\label{table:outpoison}
\scalebox{0.9}{
\begin{tabular}{l|l|l|l|l|l}
\hline
\textbf{Model}                                                                                                     & \textbf{\begin{tabular}[c]{@{}l@{}}Overall Test \\ Accuracy\end{tabular}} & \textbf{\begin{tabular}[c]{@{}l@{}}`Stop' signs \\ misclassified \\ as `Speed Limit 80km/h'\end{tabular}} & \textbf{\begin{tabular}[c]{@{}l@{}}`Speed Limit 80km/h' \\ signs misclassified \\ as `Stop'\end{tabular}} & \textbf{\begin{tabular}[c]{@{}l@{}}Bit Error Ratio\\ of retrieved bits\end{tabular}} & \textbf{\begin{tabular}[c]{@{}l@{}}Integrity\\ verified\end{tabular}} \\ \hline\hline
ResNet18 (M)                                                                                                       & 98.60\%                                                                   & 0.00\%                                                                                                    & 0.00\%                                                                                                    & -                                                                                    &                                                                       \\ \hline
\begin{tabular}[c]{@{}l@{}}ResNet18 \\ + DeepiSign-G ($\tilde{M}$)\end{tabular}                       & 98.60\%                                                                   & 0.00\%                                                                                                    & 0.00\%                                                                                                    & 0.00\%                                                                               & True                                                                  \\ \hline
\begin{tabular}[c]{@{}l@{}}\\$\tilde{(M)}$ after 1 batch\\ of output poisoned retraining\end{tabular}   & 98.58\%                                                                   & 0.00\%                                                                                                    & 0.00\%                                                                                                    & 40.56\%                                                                              & False                                                                 \\ \hline
\begin{tabular}[c]{@{}l@{}}\\$\tilde{(M)}$ after 5 epochs \\ of output poisoned retraining\end{tabular} & 91.13\%                                                                   & 99.26\%                                                                                                   & 98.57\%                                                                                                   & 50.37\%                                                                              & False                                                                 \\ \hline\hline
\end{tabular}}

\end{table*}

\subsection{Direct Targeted Tampering}
\subheading{Attack}
In this attack, we demonstrate that integrity breaches can be detected when modifications are made to a highly localized portion of the model parameters, contrasting with attacks that require adjustments to a large number of parameters. This type of attack can significantly impact model behavior, especially if the modified weights are in critical locations, such as the output layer.

\subheading{Our Implementation}
For this demonstration, we targeted a ResNet18 model's final fully connected output layer. Specifically, we zeroed the 512 adjacent weights leading to the ``Stop'' class, resulting in the model misclassifying ``Stop'' sign inputs. We used DeepiSign-G to embed hidden data into the model trained on the original dataset. The model was then modified by zeroing the 512 adjacent weights leading to the ``Stop'' class in the final fully connected output layer.

\subheading{Results}
The results of this attack are summarized in Table \ref{table:targeted}. Despite the model parameters being adjacent and distributed across multiple transform blocks, the modification was spread across many coefficients. This widespread corruption led to the detection of an integrity breach during verification, highlighting the effectiveness of DeepiSign-G.



\begin{table}[h!]
\centering
\caption{Results of a targeted modification attack on a classifier trained on the GTSRB dataset. The weights and biases to the `Stop' class are zeroed, and the integrity breach detection by DeepiSign-G is examined.}
\label{table:targeted}
\begin{tabular}{l|l|l}
\hline
\textbf{Model}                                                                                           & \begin{tabular}[c]{@{}l@{}}Bit Error Ratio \\ of Retrieved Data\end{tabular} & \begin{tabular}[c]{@{}l@{}}Integrity \\ verified\end{tabular} \\ \hline\hline
\begin{tabular}[c]{@{}l@{}}ResNet18 +\\ DeepiSign-G ($\tilde{M}$)\end{tabular}              & 0.00\%                                                                       & True                                                          \\ \hline
\begin{tabular}[c]{@{}l@{}}\\$\tilde{(M)}$ with weights\\ to `Stop' class zeroed\end{tabular} & 4.34\%                                                                       & False                                                         \\ \hline\hline
\end{tabular}

\end{table}

\begin{table*}[h!]
\centering
\caption{Results of the arbitrary modification experiment for a ResNet18 and DenseNet161 model. Different fractions of model parameters are modified by adding Gaussian noise, and DeepiSign-G detection is examined. The Bit Error Ratio provides insight into the retrieved data's corruption level.}
\label{table:arbmod}
\begin{tabular}{l|l|l|l|l}
\hline
                                                                                                       & \multicolumn{2}{l|}{\textbf{Resnet18}}                                                                                                                        & \multicolumn{2}{l}{\textbf{Densenet161}}                                                                                                                    \\ \hline
\textbf{\begin{tabular}[c]{@{}l@{}}Percentage of parameters \\ with Gaussian noise added\end{tabular}} & \textbf{\begin{tabular}[c]{@{}l@{}}Bit Error Ratio \\ of retrieved data\end{tabular}} & \textbf{\begin{tabular}[c]{@{}l@{}}Integrity\\ verified\end{tabular}} & \textbf{\begin{tabular}[c]{@{}l@{}}Bit Error Ratio\\ of retrieved data\end{tabular}} & \textbf{\begin{tabular}[c]{@{}l@{}}Integrity\\ verified\end{tabular}} \\ \hline \hline
0\%                                                                                                    & 0.00\%                                                                                & True                                                                  & 0.00\%                                                                               & True                                                                  \\ \hline
0.00001\%                                                                                              & 0.0099\%                                                                              & False                                                                 & 0.0084\%                                                                             & False                                                                 \\ \hline
0.0001\%                                                                                               & 0.11\%                                                                                & False                                                                 & 0.095\%                                                                              & False                                                                 \\ \hline
0.001\%                                                                                                & 0.93\%                                                                                & False                                                                 & 8.99\%                                                                               & False                                                                 \\ \hline
0.01\%                                                                                                 & 9.32\%                                                                                & False                                                                 & 43.13\%                                                                              & False                                                                 \\ \hline
0.1\%                                                                                                  & 43.82\%                                                                               & False                                                                 & 49.79\%                                                                              & False                                                                 \\ \hline
1\%                                                                                                    & 49.98\%                                                                               & False                                                                 & 50.26\%                                                                              & False                                                                 \\ \hline \hline
\end{tabular}

\end{table*}
\subsection{Arbitrary Tampering}
\subheading{Attack}
In this attack, we explore the detection of minor modifications to model parameters that would not significantly affect model behavior. This scenario is similar to the arbitrary modification attack described in \cite{he2018verideep}. We add Gaussian noise with mean 0 and unit standard deviation to small subsets of the parameters to see if the proposed defense can detect these minute modifications.

\subheading{Our Implementation}
Using DeepiSign-G, we embedded hidden data into the model trained on the original dataset. We then introduced Gaussian noise with mean 0 and unit standard deviation to small subsets of the parameters.

\subheading{Results}
The results of this attack are summarized in Table \ref{table:arbmod}. Despite the minor nature of the modifications, where only a few parameters were slightly adjusted, DeepiSign-G efficiently detected even these subtle integrity breaches.


\section{Discussion}\label{sec:discussion}

\subheading{Meeting the Design Requirement} 
In developing DeepiSign-G, we rigorously adhered to the six key design requirements to ensure its effectiveness and practicality in securing deep learning models.
(1) Integrity: Our approach successfully detects any unauthorized modifications to the model, as evidenced by the results of various attack scenarios.
(2) Tracking: DeepiSign-G embeds a watermark that includes all necessary metadata, enabling straightforward verification of the model's authenticity.
(3) Capacity: Unlike traditional watermarking methods limited to embedding small, constant messages to avoid distorting model parameters, DeepiSign-G can embed information in a vast majority of the  Walsh-Hadamard coefficients. For instance, in a model like ResNet18 with 11 million tunable parameters, DeepiSign can embed data in approximately 9.9 million coefficients, excluding only a small proportion of the high-frequency components.
(4) Accuracy/Invisibility: Our experiments demonstrate that DeepiSign-G has minimal impact on model accuracy, offering two significant advantages. First, the model remains usable even after watermarking, eliminating the need for watermark removal. Second, it is challenging for adversaries, even those with access to the model's open-source version, to discern whether the model has been watermarked.
(5) Confidentiality: DeepiSign-G is designed with robust security in mind, leveraging the AES256 security key to protect against unauthorized data retrieval.
(6) Our experiments confirm that DeepiSign-G is architecture-agnostic, seamlessly integrating with various model architectures, including CNNs and RNNs. Unlike previous works, such as \cite{He_2019_CVPR} and \cite{abuadbba2021deepisign}, which are specifically designed for computer vision tasks and architectures, DeepiSign-G does not suffer from inapplicability to RNN-type architectures. It demonstrates efficacy for both CNN and RNN tempering threat models.

\subheading{Comparison to previous work} 
The most closely related work is~\cite{abuadbba2021deepisign}, where a wavelet transform was employed to embed metadata within CNN models. However, DeepiSign-G offers several key advantages over this approach. Firstly, it boasts significantly lower computational complexity. The process of producing the multidimensional sub-bands wavelets tree in~\cite{abuadbba2021deepisign} is computationally expensive, with quadratic complexity in terms of both time and operations~\cite{akansu2001multiresolution}. In contrast, DeepiSign-G relies on a much lighter and faster transformation technique, namely the fast Walsh–Hadamard transform, which exhibits (i.e., linearithmic complexity n log n) in terms of time and requires operations based on additions and subtractions~\cite{fino1976unified}.

Secondly, DeepiSign-G offers a higher embedding capacity. In~\cite{abuadbba2021deepisign}, 50\% of transformed coefficients cannot be modified due to constraints within the wavelet tree, limiting the hiding capacity and overall security. In contrast, DeepiSign-G leverages the flexibility of the fast Walsh–Hadamard transform, enabling up to 90\% of the coefficients to be utilized in the hiding process. This significantly enhances the embedding capacity and security of the approach, making it a more robust choice for secure data embedding in DNN models.

\subsection{Related Work}\label{sec:relatedwork}

This section provides an overview of related works on attacks and defenses targeting DNN model integrity.

\subheading{Poisoning Attacks} Several techniques aim to compromise DNN integrity by inserting backdoors. Gu et al. \cite{gu2017badnets} introduced a poisoning attack in BadNets, creating a poisoned model through retraining with a tainted dataset. The backdoor remains active even after transfer learning to a new model. Liu et al. \cite{rw:liu2017trojaning} improved this attack by tampering with a subset of weights to inject a backdoor. Chen et al. \cite{rw:chen2017targeted} proposed an attack where the attacker reengineers the model from scratch and trains it with a poisoned dataset.

\subheading{Poisoning Defenses} Defense against backdoor attacks is actively researched. Liu et al. \cite{rw:liu2017neural} proposed three defense mechanisms, including anomaly detection in training data, retraining the model to remove backdoors, and preprocessing input data to remove triggers. He et al. \cite{rw:he2018verideep} introduced a defense technique using sensitive input samples to spot changes in hidden weights and produce different outputs. However, these defenses will not fully protect against sophisticated attacks.

\subheading{Cryptography Methods} One approach is to use cryptographic methods like digital signatures and authentication codes to safeguard the integrity and authenticity of CNN models. However, managing and distributing these signatures securely presents challenges. If a signature is lost or altered, it becomes challenging to ascertain if the model has been compromised. To mitigate this risk, protecting the signature itself may be required, potentially necessitating the establishment of additional infrastructure such as certificate authorities. Additionally, each new DNN model requires its own signature, resulting in the need for secure storage of multiple signatures alongside all metadata, which can be burdensome in environments with limited security measures.

\subheading{Adversarial Samples} Adversarial samples are crafted to evade a trained DNN model at testing time without poisoning the model itself. Attacks like the fast gradient sign method \cite{rw:goodfellow2014explaining}, basic iterative method \cite{rw:kurakin2016adversarial}, and defenses like feature squeezing \cite{rw:featuresqueezing2018} are active areas of research in this domain. This stream of work is very promising in a black-box setup to determine if the incoming input is benign or adversarial. However, they cannot find out if the integrity of
DNN model itself is maintained or violated by poisoning
attacks.

\subheading{Watermarking for IP Protection} Watermarking is used to protect the IP of DNN models. Techniques like embedding a watermark into deep layers \cite{rw:uchida2017embedding} and using 1-bit watermarking \cite{rw:merrer2017adversarial} have been proposed. However, these approaches focus on claiming ownership rather than protecting model integrity against poisoning attacks.

{\textbf{Watermarking for Integrity: }} Ensuring the integrity and authenticity of DNN models is critical. While IP watermarking focuses on ownership, methods for tracking model integrity are lacking. Our prior work~\cite{abuadbba2021deepisign} introduced the first fragile watermark for safeguarding the integrity and authenticity of models in computer vision tasks. He et al.~\cite{He_2019_CVPR} also investigated using generated sensitive samples to assess computer vision model integrity. However, both studies leave unanswered questions: \textit{i) are these methods applicable to other domains? ii) If so, how efficient are they?} Our investigation suggests that these methods are tailored to computer vision, limiting their effectiveness in other domains, such as Natural language processing (RNN). This underscores the need for more generic and secure watermarking schemes independent of specific model architectures to accommodate a broader range of neural networks.

\section{Conclusion}\label{sec:sec:conclusion}

We introduce DeepiSign-G, a mechanism for safeguarding the integrity and authenticity of deep learning models. DeepiSign-G hides data in Walsh-Hadamard coefficients, inspired by frequency-space watermarking techniques from the image and multimedia domain. This approach allows information to be stored in deep learning models without compromising their performance. We address two major challenges of the DNN architecture, namely generality and accuracy, and we ensure the mechanism is applicable across diverse models.

DeepiSign-G provides a self-contained mechanism for verifying model integrity (by checking the retrieved data's hash) and authenticity (using a secure key in embedding and retrieval). Additionally, it offers the potential to track and secure model metadata within the model itself. We demonstrate the detection performance of DeepiSign-G in detecting various integrity breaches, including a trojan attack on a text sentiment classifier.

\bibliographystyle{unsrt}  
\bibliography{references}

\end{document}